\documentstyle[preprint,aps]{revtex}
\begin{document}
\draft 
\title{\Large  Four Fermion Field Theories and the  Chern-Simons Field: 
A Renormalization Group Study}
\author{V. S. Alves\cite{byline}, M. Gomes, S. V. L. Pinheiro\cite{byline}
and  A. J. da Silva}
  \address{Instituto de F\'\i sica, USP\\
 C.P. 66318 - 05315-970, S\~ao Paulo - SP, Brazil}
\date{1999}

\maketitle

\begin{abstract}
  In (2+1) dimensions, we consider the model of a $N$ flavor,
  two-component fermionic field interacting through  a Chern-Simons
  field besides a four fermion self-interaction which consists of a
  linear combination of the Gross-Neveu and Thirring like terms. The
  four fermion interaction is not perturbatively renormalizable and
  the model is taken as an effective field theory in the region of low
  momenta. Using Zimmerman procedure for reducing coupling constants,
  it is verified that, for small values of the Chern-Simons parameter,
  the origin is an infrared stable fixed point but changes to
  ultraviolet stable as $\alpha$ becomes bigger than a critical
  $\alpha_c$. Composite operators are also analyzed and it is shown
  that a specific four fermion interaction has an improved
  ultraviolet behavior as $N$ increases.
\end{abstract}
\
Fermionic quartic interactions have been very important for the
clarification of conceptual aspects as well as for the applications of
Quantum Field Theory. Illustrative examples of such dual role are
provided by the Thirring and Nambu-Jona Lasinio models. However,
perturbative studies of the models have been hampered by the fact that
only in two dimensions they are renormalizable. If the number of
flavors is high enough, a better ultraviolet behavior is achieved in
the context of the $1/N$ expansion which turns out to be
renormalizable up to $4 - \epsilon$
dimensions\cite{Gross,Gomes1,Krasnikov}. Various studies have been
performed using such scheme\cite{1/N}.

On the other hand, for small $N$ we may consider the models as
effective field theories\cite{Weinberg}, reliable at low energies, as
it has indeed been done in their phenomenological
applications\cite{Bardeen}.  Besides that, recent studies
\cite{Chen,Alves} pointed out that in 2+1 dimensions yet another
complementary direction would be available.  Through the interaction
with a Chern Simons (CS) field \cite{Jackiw} fermionic fields could
change their operator dimension in such way to improve the ultraviolet
behavior of the perturbative expansion.  In\cite{Alves} this
conjecture was investigated for the case of Gross-Neveu model coupled
to a Chern Simons field and considering $N=1$.  Although the
improvement does happen for the basic field, we found that quartic
composite operators do not share this property.  This means that the
behavior of these operators is not affected by the CS field.
Nevertheless, from the characteristics of the $1/N$ expansion and also
from the non-perturbative investigations based in the Schwinger Dyson
equation we may expect the existence of relevant four fermion
interactions when $N\not=1$. In fact, nonperturbative studies point
towards the existence of critical values of $N$ where mass generation
occurs and basic properties of the theories are drastically
changed\cite{Appelquist,Gomes1}.  It is therefore reasonable to expect
substantial changes in these theories as $N$ increases, even at the
perturbative level.

In this short communication, pursuing the work of \cite{Alves}, we
will present some results on four fermion theories coupled to a CS
field when $N$ is small but $\not= 1$. The basic field $\psi$,
belonging to the two dimensional representation of the Lorentz group
has now both Lorentz and SU(N) indices which we will sometimes
indicate by Greek and Latin letters, respectively.  Our first
observation is that, due to the Fierz identity\cite{Gomes},
\begin{eqnarray}
(1+\frac{2}{N}) (\bar \psi \psi)^2 + (\bar\psi \gamma^\mu \psi)(\bar\psi 
\gamma_\mu \psi) + (\bar \psi \lambda^a \psi)(\bar \psi \lambda^a \psi) &=&0,\\
2(2 + \frac 1N)(\bar \psi \psi)^2 +\frac{2}{N}(\bar\psi \gamma^\mu \psi)
(\bar\psi \gamma_\mu \psi) + (\bar \psi \gamma^\mu\lambda^a \psi)
(\bar \psi \gamma_\mu\lambda^a \psi) & = & 0,
\end{eqnarray}
where $\lambda^a, \,\, a=1,\ldots,N^2 -1$ are the generators of SU(N), there
are only two independent, Lorentz and SU(N)  scalar quartic self-interactions.
Therefore, without loosing generality, we may restrict our study to
the theory described by the Lagrangian,
\begin{eqnarray}
{\cal L}& =& \frac{1}{2\pi\alpha}
\varepsilon^{\mu\nu\alpha}\,\partial_{\mu} A_\nu \,A_{\alpha}
+\bar{\psi}(i \not \! \partial-m) \psi+ \bar \psi\gamma^\mu \psi A_\mu
- G_1(\bar \psi\psi)(\bar \psi\psi)\nonumber \\
&\phantom a& -G_2 (\bar\psi \gamma^\mu \psi)
(\bar\psi \gamma_\mu \psi) + \frac{1}{2\xi}(\partial_\mu A^\mu)^2. 
\label {1}
\end{eqnarray}
Actually, evading possible infrared divergences, throughout this paper
we will work in a Landau gauge obtained by formally letting
$\xi \rightarrow 0$. As the canonical dimension of $\psi$ is
one, both $G_1$ and $G_2$ have dimension -1 in mass unity.  The model
is therefore nonrenormalizable, the degree of superficial divergence
of a generic graph with $N_A$ and $N_F$ bosonic and fermionic external
lines, and with $V_1$ and $V_2$ Gross-Neveu and Thirring like
vertices, being equal to
\begin{equation}
d(\gamma)= 3 - N_A - N_F +V_1 + V_2.
\end{equation}
To validate our calculations we shall treat (\ref{1}) as an effective
theory, suppressing the high momenta contributions to the Feynman
amplitudes.  This is conveniently done by introducing a dimensional
parameter $\Lambda$ through the definitions $G_1=g_1/\Lambda$ and
$G_2=g_2/\Lambda$ and restricting the calculation by requiring that
$p\ll \Lambda$.  In this implementation $g_1$ and $g_2$ must then be
considered as the perturbative couplings.

To regulate Feynman integrals, we use the following ``dimensional 
regularization'' recipe.  Initially, the algebra of the Dirac matrices and
contractions of the $\varepsilon$ Levi-Civita symbols are performed in $2+1$
dimensions using
\begin{equation}
\gamma^\mu \gamma^\nu = g^{\mu\nu} - i \varepsilon^{\mu\nu\rho} \gamma_\rho ,
\end{equation}
and
\begin{equation}
\varepsilon^{\mu\nu\rho} \varepsilon_{\rho\sigma\lambda}= \delta^{\mu}_{\sigma}
\delta^{\nu}_{\lambda}- \delta^{\mu}_{\lambda} \delta^{\nu}_{\sigma},
\end{equation}
\noindent
After this step, the integrals are promoted to $d$ dimensions and
carried out accordingly the usual rules \cite{Collins}. Singularities
appear as poles at $d=3-\epsilon$ which should then be removed. To
this end, to each loop integral we incorporate the factor
$\mu^{\epsilon}$ where the massive parameter $\mu$ plays the role of
the renormalization point.  The renormalized amplitude is given by the
constant term (i.e., the $\epsilon$--independent one) of the Laurent
expansion of the resulting expression. This ``dimensional
regularization'' method does not require an extension of the
Levi-Civita symbol outside $2+1$ dimensions and thus is very
convenient for practical calculations.  One should be aware that
slight modifications of these rules may change the finite part (for
example, using $\gamma^\mu \gamma^\alpha \gamma_\mu$ = 2--d instead of
-1) of the outcome. However, our results will not be affected since we
will be dealing only with the simple pole part of the amplitudes
(double poles only appear at higher orders, i. e., in the computation
of graphs with three or more loops).  Actually, with the mentioned
restrictions the method has been applied and tested in variety of
problems in $2+1$ dimensions\cite{Chen,Avdeev}.

The vertex functions so defined approximately satisfy a renormalization
group equation,
\begin{equation}\label{2}
[\Lambda \frac{\partial\phantom {a}}{\partial \Lambda} + \mu 
\frac{\partial\phantom {a}}{\partial \mu}+
\beta_1 \frac{\partial\phantom{a}}{\partial g_1}+
\beta_2 \frac{\partial\phantom{a}}{\partial g_2}-\gamma N_F ]\Gamma^{(N)}(p_1, 
\ldots p_N)\approx 0 ,
\end{equation}

\noindent
where, as a consequence of the Coleman--Hill theorem \cite{Coleman}, a
term proportional to the derivative of the $\alpha$ parameter is
absent. The coefficients $\gamma$ and $\beta_i$ can be calculated by
replacing the two and four point function into (\ref{2}). 

To fix $\gamma$ notice that, up to 2 loops, only graphs which are
second order in $\alpha$ may contribute to the wave function
renormalization (i.e., linearly divergent graphs with 2 external
fermionic lines).  There are 3 graphs (the same as in Fig 2 of
\cite{Alves}) and a direct computation gives
\begin{equation}\label{3}
\gamma= -\frac{N+1}{24} \alpha^2.
\end{equation}

\noindent
Notice that for $N=1$ this result agrees with \cite{Alves}, as it
should.  

Analogously, $\beta_1$ and $\beta_2$ can be determined from
the momentum independent residues in the four point vertex functions.
In this calculation, it should be observed that the $\mu$ dependence
of pole part arises through the expansion of the $\mu^\epsilon=1 +
\epsilon \ln \mu + O(\epsilon^2)$ factors, introduced for each loop
momentum integral.  We denote the Fourier transform of $\langle 0|{\rm
  T} \psi_{\alpha_1 a_1}(x_1) \psi_{\alpha_2 a_2}(x_2)\bar
\psi_{\alpha_3 a_3}(y_1)\bar \psi_{\alpha_4 a_4}(y_2)|0\rangle$ by
$\Gamma^{(4)}_{\alpha_1 a_1, \alpha_2 a_2;\alpha_3 a_3,\alpha_4 a_4}$,
where Lorentz and $SU(N)$ indices are represented by greek and latin
letters, respectively.
We found that, up to third order in $g_1,\,\, g_2$ and
$\alpha$, the $\mu$ dependence of the 4 point function is given by
\begin{eqnarray}
&&\mbox{$\mu$ dependent part of}\,\,\, \Gamma^{(4)}_{\alpha_1 a_1,
\alpha_2 a_2;\alpha_3 
a_3,\alpha_4 a_4}(p_i=0)\nonumber \\
&=&-2 i\alpha^2 \ln \mu \Bigl[(\frac{g_1}{\Lambda} 
(\frac72+ 3 N )+ \frac{g_2}{\Lambda} (9+ 3N)) (\Delta \otimes \Delta) +
 (\frac{g_1}{\Lambda}  (\frac52+\frac{N}3)\Bigr.\nonumber\\
 &\phantom a&\Bigl. -\frac{g_2}{\Lambda} 
(\frac23+\frac{N}3)) (\Gamma\otimes\Gamma)\Bigr ]_{\alpha_1 a_1,\alpha_2 a_2,
\alpha_3 a_3,\alpha_4 a_4}.
\end{eqnarray} 
\noindent
where we adopted the simplified notation
\begin{eqnarray} 
(\Delta\otimes \Delta)_{\alpha_1 a_1,\alpha_2 a_2;\alpha_3 a_3,\alpha_4 a_4} 
&=&
\delta_{\alpha_1\alpha_3}\delta_{\alpha_2\alpha_4}\delta_{a_1a_3} 
\delta_{a_2 a_4}-\delta_{\alpha_1\alpha_4}\delta_{\alpha_2\alpha_3}
\delta_{a_1a_4} \delta_{a_2 a_3}\\
(\Gamma\otimes \Gamma)_{\alpha_1 a_1,\alpha_2 a_2;\alpha_3 a_3,\alpha_4 a_4} 
&=&\gamma^{\mu}_{\alpha_1\alpha_3}\gamma_{\mu \alpha_2\alpha_4}\delta_{a_1 
a_3}\delta_{a_2 a_4}-\gamma^{\mu}_{\alpha_1\alpha_4}\gamma_{\mu \alpha_2
\alpha_3} \delta_{a_1 a_4}\delta_{a_2 a_3}
\end{eqnarray}
for the Gross-Neveu and Thirring vertices, respectively.

Replacing the above expression into (\ref{2}), using (\ref{3}) and
equating to zero the coefficients of the Gross-Neveu and Thirring
vertices we determine $\beta_1$ and $\beta_2$ to be
\begin{eqnarray}
\beta_1&=& g_1 - \frac{43+37N}{6}\, g_1 \alpha^2 - 2(9+ 3N) g_2 \alpha^2 ,\\
\beta_2&=& g_2 - (5+\frac{2N}3) g_1 \alpha^2 + \frac12(\frac73+N) g_2 \alpha^2.
\end{eqnarray}

\noindent
Since the Gross Neveu and Thirring interactions were taken as
independent, these expressions are valid only if $N>1$.  They show
that the renormalization group fixed points, defined through the
vanishing of $\beta_1$ and $\beta_2$, will require
\begin{equation}
\alpha^2 =\alpha^{2}_{c}=\frac{6[-17N - 18 + (\Theta)^{1/2}]}
{(3541+1900 N +255N^2)},
\end{equation}
with $\Theta = 3865 + 2512 N + 544 N^2$.
However, to better understand the nature of this result it is
convenient to use the  systematic procedure devised by Zimmermann
\cite{Zimmermann} which allows us to consider just one constant, $g_1$
let us say, as independent.  The other coupling is then fixed as to
have just one $\beta$ function in the renormalization group equation.
Such scheme has been applied in a variety of circumstances, including
cases of nonrenormalizable models treated as effective
theories\cite{Toro}.  We thus suppose that $g_2= \rho_0 g_1$ where
$\rho_0$ is a constant such that $\beta_2 = \rho_0 \beta_1$, which
gives 
\begin{equation}
\rho_0 =\frac{-25 - 20 N+(\Theta)^{1/2}}{36(3+N)}\,.  
\end{equation}
In this situation,
\begin{equation}
\beta_1 = g_1 \{1 - \alpha^2 [\frac{43+37N}{6} + (18+ 6 N)\rho_0]\}\, .
\end{equation}
From this equation we conclude that the origin is an infrared fixed point
stable or unstable accordingly $\alpha < \alpha_c$ or $\alpha>\alpha_c$. At $\alpha=\alpha_c$, $\beta_1=0$ and the
theory is approximately scale invariant. 

We want now to go back to the question posed at the beginning of this
paper, namely, if $N>1$ does the coupling with the CS field improve
the ultraviolet behavior of quartic operators?  If this were the case
one could use this quartic interaction to perturb the model of
fermionic particles interacting just through a CS field.  We thus
consider $g_1=g_2=0$ and study the renormalization behavior of
integrated operators of canonical dimension 4. Specifically, we define 
(symbolically)
\begin{equation}
\Delta_1 = \int d^3x \overline \psi D^2\psi\, , \qquad \Delta_2=\int d^3x
(\overline \psi \psi)^2 \quad \mbox{and} \quad \Delta_3=\int d^3x (\overline \psi \gamma^\mu \psi) (\overline \psi \gamma_\mu \psi)\, ,\label{4}
\end{equation}
where $D^2=D_\mu D^\mu$ and $D_\mu = \partial_\mu - -i \sqrt \alpha
A_\mu$ is the covariant derivative.  The renormalized integrated
operators are obtained by removing poles so that, up to second order
in $\alpha$, in momentum space the renormalized amplitude with the
insertion of the operator $\Delta_i$ is
\begin{equation}
\Gamma_{\Delta_i} = (1-\tau) I_{\Delta_i}= (\delta_{ij} + z_{ij})\Gamma^{\prime}_{\Delta_i}
\end{equation}
where $I_{\Delta_i}$ is the dimensionally regularized integral,
$\Gamma_{\Delta_i}^{\prime}$ is the $\mu$-independent part of $\Gamma_{\Delta_i}$ and the
matrix $z$ is given by
\begin{equation}
 z=2\alpha^2\ln[\mu] \left (\begin{array}{lrr}-
\frac{1}{3}&0&0\\
0&7/2+ 3N&5/2+N/3\\
(1+N/4)\frac{1}{6\pi\alpha}& 9 + 3N & -2/3 -N/3 \, .
\end{array}\right )
\end{equation}
With this understanding we may write 
$\Delta_{iR} = (\delta_{ij} + z_{ij})\Delta_{j}^\prime$, where $\Delta^{\prime}_{j}$ is the finite part corresponding to $\Gamma^{\prime}_{\Delta i}$

Although the operators $\Delta_i$ in (\ref{4}) are not multiplicative renormalized we can find
new operators having such property by taking adequate linear combinations
$\bar \Delta_{i} =C_{ij} \Delta_{j}$. The new renormalized operators are
then linear combinations od the old ones, $\bar \Delta_{iR} =C_{ij} \Delta_{jR}$.
 
The specific form of the
matrix $C$ is not actually relevant but it is such that
$\bar \Delta_{iR} = (\delta_{ij}+Z_{ij}) \bar \Delta_{j}$ where $Z$ is a diagonal
matrix. We found 

\begin{equation}
Z=2 \alpha^2\, \ln\mu\, \mbox{Diagonal}\Bigl ( -1/3\, ,\,\, \frac{1}{12}( -\sqrt\Theta + 17 + 16 N), \,\,\frac{1}{12}(\sqrt\Theta + 17 + 16 N)\Bigr)\, .
\end{equation}
 We are now in a position to calculate
the anomalous dimension for these operators.  Indeed, from the above
results and  noticing that they
satisfy
\begin{equation}
(\mu\frac{\partial\phantom a}{\partial \mu} -\gamma N_F +  \gamma_{\bar
\Delta_{iR}} ) \Gamma_{\bar \Delta_{iR}}^{(N_F)}=0 
\end{equation}
we arrive at
\begin{equation}
 \gamma_{\bar \Delta_{1R}} = \frac{7-N}{12} \alpha^2,\qquad
\gamma_{\bar\Delta_{2R}} = \frac{1}{6}(\sqrt \Theta -17N -18 )\alpha^2,
\end{equation}
and
\begin{equation}
\gamma_{\bar \Delta_{3R}} = -\frac{1}{6}(\sqrt \Theta +17N +18 )\alpha^2
= -\frac{\alpha^2}{\alpha^{2}_{c}}\, .
\end{equation}

Thus, in the infrared stable region, $\alpha< \alpha_c$, there are two
operators ($\bar \Delta_{1R}$ and $\bar \Delta_{3R}$) whose dimensions
decrease with $N$. The anomalous dimension $\gamma_{\bar \Delta_{1R}}$
has a very small variation implying that the ultraviolet behavior of
the corresponding operator is not improved in a meagninfull way. The
situation is much better concerning the second operator.  By
conveniently choosing $\alpha$ near $\alpha_c$, the operator dimension
of $\bar \Delta_{3R}$ may become near 3 as we want and, for all
practical purposes, the interaction behaves like a renormalizable one.  This
operator is therefore a natural candidate for implementing a
consistent perturbation scheme around the conformal invariant theory
of fermions interacting through a Chern--Simons field. Of
course, higher order corrections may modify the above results. Thus,
increasing the parameter $\Lambda$ will require the inclusion of new
interactions and in principle new couplings will be needed. However,
we may conjecture that in this phase, following Zimmermann's
procedure, it will also be possible to fix the new couplings as
definite functions of just one four fermion coupling.

This work was supported in part by Conselho Nacional de
Desenvolvimento Cient\'\i fico e Tecnol\'ogico (CNPq) and Coordena\c
c\~ao de Aperfei\c coamento de Pessoal de N\'{\i}vel Superior (Capes).


\begin{thebibliography}{99}
\bibitem[*]{byline} On leave of absence from Universidade Federal do Par\'a.

\bibitem{Gross} D. Gross in Methods in Field Theory, Les Houches 1975,
  R. Balian and J. Zinn-Justin eds., North-Holland, Amsterdam, (1976);
  G. Parisi, Nucl. Phys. B100, 368 (1975); B. Rosenstein, B. J. Warr and 
  S. H. Park, Phys. Rev. Lett 62 1433, (1989).


\bibitem{Gomes1} M. Gomes, R. S.Mendes, R. F. Ribeiro, and A. J. da 
Silva, Phys. Rev. D {\bf 43}, 3516 (1991). 

\bibitem{Krasnikov} N. V. Krasnikov and A. B. Kyatkin, Mod.  Phys. Lett. 
{\bf A6}, 1315 (1991).

\bibitem {1/N} B. Rosenstein, B. J. Warr and S.H. Park, Phys. Rev. D 
{\bf 39} 3088, 1989   L. Del Debbio, S. J. Hands, J. C. Mehegan,  Nucl. 
Phys. B502 (1997) 269; S. Hands,  ``Fixed Point Four-Fermi Theories'', 
hep-lat/9706018; J. Zinn-Justin, Nucl. Phys. {\bf B367}, 105 (1991).

\bibitem{Weinberg} S. Weinberg, ``The Quantum Theory of Fields'', Cambridge
  University Press, 1995; G. P. Lepage proceedings of TASI-89, 1989;
G. P. Lepage {\it et al.}, Phys. Rev. D {\bf 46}, 4052 (1992);
M. Luke, A. V. Manohar, Phys. Rev. D{\bf 55}, 4129 (1997).   

\bibitem {Bardeen} S. P. Klevansky, Rev. Mod. Phys. {\bf 64}, 649 (1992);
W. A. Bardeen, C. T. Hill and M. Lindner, Phys. Rev. D {\bf 41},
(1990) 2197; M. Atance and L. Cort\'es, Phys. Rev. D {\bf 54}, 4973 (1996).

\bibitem{Chen} W. Chen and M. Li, Phys.  Rev. Lett. 70, 884 (1993); W.
  Chen, Nucl. Phys. B435, 689 (1995).

\bibitem{Alves} V. S. Alves, M. Gomes, S. V. L. Pinheiro and  A. J. da Silva,
Phys. Rev. D {\bf 59}, 045002 (1999).

\bibitem {Jackiw} S. Deser, R. Jackiw and S. Templeton, Phys. Rev. Lett. 
{\bf 48}, 975 (1982); Ann. Phys. (N.Y.) {\bf 140}, 372 (1982); J. F. 
Schonfeld, Nucl. Phys. B185, 157 (1981).

\bibitem {Appelquist} T. Appelquist, D. Nash, and 
L. C. R. Wijewardhana, Phys. Rev. Lett. {\bf 60} 2575 (1988).

\bibitem{Gomes}  M. Gomes, V. O. Rivelles and A.  J. da Silva, Phys. Rev. D
{\bf 41}, 1363 (1990).

\bibitem{Collins} J. C. Collins, ``Renormalization'', Cambridge University
Press, 1984.

\bibitem{Avdeev} L. V. Avdeev, G. V. Grigoryev and D.I.Kazokov, Nucl. Phys.
{\bf B382}, 561 (1992); G. W. Semenoff, P. Sodano and Yong-Shi Wu,
Phys. Rev. Lett. {\bf 62}, 715 (1989); W. Chen, G.W. Semenoff and Y.S.Wu,
Phys Rev. D{\bf 46}, 5521 (1992).
\bibitem{Coleman} S. Coleman and B. Hill, Phys. Lett. {\bf 159B}, 184 (1985).

\bibitem{Zimmermann} W. Zimmermann {\bf 97} 211 (1985);  R. Oehme and W. Zimmermann, Comm. Math. Phys. {\bf 97}, 569 (1985).

\bibitem{Toro} L. A. W. Toro, Z. Phys. C {\bf 56} 635 (1992);  M. Atance and L. Cort\'es, Phys. Rev. D {\bf 56} 3611 (1997). 
\end{thebibliography}
\end{document}